\documentclass[english,twocolumn]{IEEEtran}
\usepackage{graphicx, footnote, amssymb, mathtools}
 \usepackage{graphicx}
\usepackage{amsmath,epsfig,amssymb}
\usepackage{times,amsmath,epsfig,amssymb,graphicx,amsfonts}
\usepackage{graphicx, footnote, amssymb, mathtools}
 \usepackage{graphicx}
\usepackage{amsmath,epsfig,amssymb}
\usepackage{times,amsmath,epsfig,amssymb,graphicx,amsfonts}

\usepackage{color}
\usepackage{amsmath}
\usepackage{amsthm}
\usepackage{epsfig}
\usepackage{amsmath}
\usepackage{amsthm}
\usepackage{epsfig}
\newtheorem{theorem}{Theorem}
\newtheorem{lemma}{Lemma}
\newtheorem{corollary}{Corollary}

\newtheorem{definition}{Definition}
\newtheorem{proposition}{Proposition}
\newtheorem{note}{Note}

\begin{document}
%%%%%%%%%%%%%%%%%

%%%%%%%%%%%%%%%%%%%%%%%%%%%%%%%%%%%%%%%%%%%%%%%%%%%%%%%%%%%%%%%%%%%%
\title{On Locally Decodable Source Coding}

\author{ Ali Makhdoumi, Shao-Lun Huang,   Muriel M{\'e}dard, Yury Polyanskiy\\
Massachusetts Institute of Technology,\\
Email: \{makhdoum, shaolun,  medard, yp\}@mit.edu,
}

\maketitle

%%%%%%%%%%%%%%%%%%%%%%%%%%%%%%%%%%%%%%%%%%%%%%%%%%%%%%%%%%%%%%%%%%%%

\begin{abstract}
 Locally decodable channel codes form a special class 
 of error-correcting codes with the property that the decoder
  is able to reconstruct any bit of the input message from 
  querying only a few bits of a noisy codeword. It is well 
  known that such codes require significantly more 
  redundancy (in particular have vanishing rate) compared 
  to their non-local counterparts. In this paper, we define a dual problem, 
   i.e. locally decodable source codes (LDSC).
     We consider both almost lossless (block error) and lossy (bit error) cases.
      In almost lossless case, we show that
        optimal compression (to entropy)
         is possible with $O(\log n)$ queries to compressed string by
          the decompressor. We also show the following converse bounds:
           $1)$ linear LDSC cannot achieve 
          any rate below one, with a bounded number of queries, 
         $2)$ rate of any source coding with linear decoder (not necessarily local) in one, 
         $3)$ for $2$ queries, any code construction cannot have 
         a rate below one. In lossy case, we show that any rate above rate distortion is achievable 
          with a bounded number of queries. We also show that, rate distortion is 
          achievable with any scaling number of queries. We provide an achievability 
         bound in the finite block-length regime
          and compare it with the existing bounds in succinct data structures literature.

\end{abstract}
%%%%%%%%%%%%%%%%%%%%%%%%%%%%%%%%%%%%%%%%%%%%%%%%%%%%%%%%%%%%%%%%%%%%
\section{Introduction} \label{sec:Intro}

\subsection{Motivation}
The basic communication problem may be expressed as transmitting data source
 with the highest fidelity without exceeding a given bit
rate, or expressed as transmitting the source data using the lowest
bit rate possible while maintaining a given reproduction fidelity \cite{shannon1948mathematical}.
In either case, a fundamental trade-off is made between bit rate and
%%%
distortion/error level. Therefore, source coding is primarily characterized by 
rate and distortion/error of the code. 
However, in practical communication systems, many issues
 such as memory access
requirements (both updating memory and querying from memory)
 must be considered. In traditional compression algorithms (both 
in theory and practice) a small change
in one symbol of the source sequence leads to a large change in the
encoded sequence. Another issue is that, in order to retrieve one symbol of the source,
accessing all the encoded symbols are required. The latter issue 
is the main topic of this paper. 

 One way to resolve these issues is to place constraints on the encoder/decoder.
In particular, in order to address the issue
 of memory reading access requirement, we study a class of codes for which the decoder is local. 
 This problem appears in many applications in distributed data management.
 For instance, assume that a given source is compressed and stored on some storage cells.
%  since writing on data storage cells is generally costly, we use source coding to decrease the cell usage.
   If we use the traditional source coding, then in order to
    recover only one bit of the original source, we would need to read the entire 
    encoded data on all data storage cells. 
    %%%
       Since reading from the storage cells is generally costly, we may 
       want to design a compression scheme, which
        only need to read part of the storage cells to recover one bit of the 
        source. %, and hopefully, can achieve certain level of compression.
         Assuming a local decoder is one possible solution for this matter.
%    
%%%%%%%%%%%%%%%%%
 In another example, assume that we encode a source and then store it on some data storage cells.
  We are asked to reveal information about one symbol/coordinate of the source to some party, but, we do not want
   to reveal the information about the entire source symbols.
    If we use a conventional source coding, we may have to reveal all the encoded data.
      Thus, a honest but curious
     party may have access to the entire original source sequence.  On the other hand,
      with a local decoder, we only provide a small part of the encoded data, so that the party 
      can only recover a small part
       of original source symbols without capability of extracting information about
        the other symbols.

\subsection{Contributions}
 
 We introduce \emph{locally decodable source coding (LDSC)}: A source sequence $x_1^n$ 
 (this denotes the vector $(x_1, \dots, x_n)$)
takes values from the source alphabet $\mathcal{X}$ and is mapped into a sequence $y_1^k$ of encoded 
symbols taking values in the alphabet $\mathcal{Y}$. These symbols are then used
to generate the reproduction sequence $\hat{x}_1^n$. 
%The
%rate of the source coding is defined as the ratio of the length of the output sequence to the input sequence.
 A scheme is called
$t$-locally decodable, if for any $i=1, \dots, n$, each reproduced 
symbol $\hat{x}_i$ is a function of at most $t$ of the symbols $y_1, \dots , y_k$. 
We shall define this notion formally in Section \ref{sec:rateofLDSC}. 
 The number of queries
 to decode any source symbol, is called locality and is shown by $t$. % as  %The problem formulation is depicted in Figure \ref{fig:formulation}.
 This is different from traditional source coding, as we are restricting
  the way that $y^k$ can be mapped back to the $\hat{x}^n$ sequence. 
  Throughout this paper, $X$ denotes a random variable taking 
  values in $\mathcal{X}$, where $x$ denotes an outcome of $X$.  
  The same notation holds for other letters such as $Y$ and $\hat{X}$. Also, 
  for any subset $S\subset \{1, \dots, n\}$, $X^S$ is defined as the vector 
  $(X_i:~ i\in S)$.

We consider almost lossless source coding with local decoder. 
We 
 provide a converse bound on the rate of linear LDSC and show that,
 the rate of linear LDSC is one rather than the entropy rate.

We also consider source coding with linear decoder and provide a converse bound on the rate 
 of any code (not necessarily local). We then show the 
 rate of source coding with linear decoder is one, whereas using linear and local encoder 
 we can achieve any rate above entropy rate as it is shown in \cite{mackay1999good}. 
% whereas, our results states that  
 Moreover, we consider a general 
 encoder and $2-$local decoder ($t=2$) and show the rate of which is one.
 
  We  
 provide achievability bound on the rate with scaling number of queries. In 
 particular, with $O(\log n)$ queries, any rate above entropy rate is achievable. 
 Furthermore, we 
 consider lossy source coding with local decoder and provide 
 achievability bound on the rate with both scaling number of queries and constant number of queries. 
\\ Scaling number of queries: we show that, with any number of queries scaling with $n$, rate 
 distortion is achievable. We provide an upper bound on the rate in the  
 finite block-length regime (finite $n$). We compare our achievability bound with 
 the existing results in the data structure literature and show that, 
 our achievability bound is tighter than the existing bound in \cite{Mihai}.
 \\ Constant number of queries: we show that,
  for any given rate above rate distortion, there exists a constant, $t$, such 
  that sequences of source symbols can be compressed with the given rate and 
  then decompressed with locality $t$, without exceeding the distortion constraint.

%%%%%%%%%%%%%%%%
\subsection{Related Work}
A long line of research has addressed a similar problem from a data structure perspective.
For example, Bloom filters \cite{bloom1970space} are data
structures for storing a set in a compressed form while allowing
membership queries to be answered in constant time. The
rank/select problem \cite{pagh2001low, jacobson1989space}
and dictionary problem in the field of succinct
data structures are also examples of problems involving both
compression and the ability to recover efficiently a single
symbol of the input message. In particular, reference \cite{Mihai} provides a succinct data structure
for arithmetic coding that supports efficient recovery of source symbols.
 Moreover, reference \cite{chandar2009locally} studies both issues 
 simultaneously and 
introduce a data structure that is efficient in both updating and querying. 
In most of these works, the
 efficiency is interpreted in terms of the decoding time, whereas in this work
it is interpreted in terms of memory access requirement.
  We formulate this problem
from an information theoretic view and study the fundamental trade-offs between locality and the 
rate of source coding.

{Causal Source Coding} is a related topic: 
  the constraint on the decoder is not locality, but,
   causality \cite{neuhoff1982causal, DBLP:journals/corr/abs-1301-0079}.

%%%%%%%%%%%%%%%%%%
 {Locally decodable codes} (LDC) (\cite{yekhanin2006new}) is a counter part in the 
 error-correction world. 
%%%%%%%%%%%%%%%%%%
Another recent variation is {Locally repairable codes} (\cite{papailiopoulos2012locally}).
 
 The problem of source coding with local encoding has
    been studied in many works in both data structure and 
   information theoretic literatures. This line of research addresses the update efficiency issue.
 Varshney et al. \cite{varshney2008malleable} analyzed  continuous source codes
from an information theoretic point of view
. Also, Mossel and Montanari \cite{montanari2008smooth} have constructed
 source codes with local encoder based on nonlinear graph codes. 
Sparse linear codes have been studied by Mackay \cite{mackay1999good}, who introduced  
a class of local linear encoders. Also, Mazumdar et al. \cite{mazumdar2012update}
have considered update efficient codes, which studies channel coding problem with
local encoders.

%%%%%%%%%%%%%%%%%%
The organization of the paper is as follows. In Section \ref{sec:rateofLDSC} we give the problem formulation
 and  the converse bound on the rate of LDSC. We also provide an achievability bound in case of scaling number of queries with block-length.
 \emph{Locally decodable lossy source coding (LDLSC)} is defined in Section \ref{sec:rateofLDLSC}, where
  we provide achievability bounds on the rate of LDLSC with both constant 
  and scaling number of queries. 
We conclude the paper in Section \ref{sec:conclusion}. %Rate of LDSC is studied in section \ref{sec:rate} and rate distortion with local decoder is studied in section \ref{sec:ratedistortion}.

%%%%%%%%%%%%%%%%%%%%%%%%%%%%%%%%%%%%%%%%%%%%%%%%%%%%%%%%%%%%%%%%%%%%

\section{Locally Decodable Source Coding (LDSC)}\label{sec:rateofLDSC}
%%% Definition of Local code with fixed number of queries %%%
%%%%%

%%%%%%%%%%%%%%%%%
First, we define LDSC and the fundamental limits of it. We then 
show converse bounds on the rate of LDSC with linear encoder, linear decoder and 
general encoder-decoder with locality, $t=2$.

An almost lossless LDSC is defined as a pair, consisting of an encoder
, $f$, and a decoder, $g$, such that $f:\mathcal{X}^n\mapsto \{0,1\}^k$ and $g:\{0,1\}^k\mapsto \mathcal{X}^n$.
%where $X^n$ denotes $(X_1, \dots, X_n)$. 
The decoder is called local if each coordinate of the output is affected by a 
bounded number of input coordinates. 
Formally, Let $g_a$, for $a\in\{1, ..., n\}$, be the $a-$th component of the decoding function.
 Assume $g_a$ depends on $\mathcal{Y}^k=\{0,1\}^k$ only
through the vector $Y^{\mathcal{N}_a}=\{Y_j~:~j\in \mathcal{N}_a\}$ for some $\mathcal{N}_a \subset \{1, ..., k\}$. 
In other words, we have:
$$\text{For any } y^k \text{ and } y'^k, g_a(y^k)=g_a(y'^k) \text{ if } y^{\mathcal{N}^X_a} = y'^{\mathcal{N}_a}.$$
For any given $t$, a decoder is called $t-$local if  $|\mathcal{N}_a| \le t$ for any $a\in\{1, ..., n\}$. 
We may represent the decoder $g$, by $n$ functions: $g_a: Y^{\mathcal{N}_a}\to \mathcal{X}$ for any $1 \le a \le n$. 
%%%%%%%%%%%%%%%%%
\begin{definition}\label{Def:localdecoder}

 A $(n,k,t,\epsilon)-$LDSC is a pair, consisting of an encoder, $f:\mathcal{X}^n\mapsto \{0,1\}^k$,
  and a $t-$local decoder, $g:\{0,1\}^k\mapsto \mathcal{X}^n$, such that
 \begin{equation}\label{Eq:bounderrorLDSC}
 \mathbb{P}[g(f(X^n))\neq X^n] \le \epsilon. 
 \end{equation} 
 %%%%%%%%%%%%%%%%%%%
Let 
\begin{equation}
k^*(n,\epsilon,t)=\min\{k:~~ \exists~ (n,k,\epsilon,t)- \text{LDSC}\}.
\end{equation}
 For a given $n$, $t$, and $\epsilon$, the best rate of LDSC is given by
\begin{align}
R^*(n,\epsilon,t)=\frac{k^*(n,\epsilon,t)}{n}, 
\end{align}

\begin{align}
R^*(\epsilon,t)=\limsup_{n\to\infty} R(n,\epsilon,t),
\end{align}
and 
\begin{align}
R^*(t)=\lim_{\epsilon\to 0} R(\epsilon,t).
\end{align}
\end{definition}
\begin{note}
In this paper we assume both encoder and decoder are deterministic. 
because for a given a $(n,k,\epsilon,t)-\text{LDSC}$ with randomized encoder and decoder, 
there exists an $(n,k,\epsilon,t)-\text{LDSC}$ code with deterministic encoder and decoder:
\\ Let $M$ and $N$ be two random variables and consider randomized encoder and 
decoder $f(M)$ and $g(N)$, respectively. Equation \eqref{Eq:bounderrorLDSC} then becomes
\begin{align}\label{Eq:averrorpr}
&{P}[g(f(X^n,M),N)\neq X^n] \nonumber\\
&=\mathbb{E}[\mathbb{P}[g(f(X^n,M),N)\neq X^n]|M,N] \le \epsilon.
\end{align}
Since the expectation in \eqref{Eq:averrorpr} is less than or equal to $\epsilon$,
 there exist $m,n$ such that
\begin{equation}
\mathbb{P}[g(f(X^n,M),N)\neq X_i|M=m,N=n]\le \epsilon,\nonumber
\end{equation}
showing that $f(m)$ and $g(n)$ are our desired deterministic encoder and decoder, respectively. 
\end{note}
Next, we prove a converse bound on the rate of LDSC with linear encoder.
\subsection{Linear Encoder}
We focus on binary sources, where $\mathcal{X}=\{0,1\}$. 
We show that, using a linear encoder, the rate of LDSC is one rather than the
 entropy rate.
 
  %%%
 In order to prove the converse bound
 we use the following lemma.
%%%%%%%%%%
\begin{lemma}\label{Lem:Linearalgebra}
 Let $\mathbb{F}_2^n$ be a vector space over $\mathbb{F}_2$. Let $\mathbb{P}_X=\text{Bern}(p)$ and
 define a probability measure over $\mathbb{F}_2^n$ according to a $n-$fold 
 product of $\mathbb{P}_X$, i.e. $\mathbb{P}^n_X$. If $U$ is a $k$-
  dimensional sub-space of $\mathbb{F}_2^n$, we have 
\begin{equation}
(\max\{p,1-p\})^{n-k} \ge \mathbb{P} [U] \ge (\min\{p,1-p\})^{n-k}.
\end{equation}
\end{lemma}
\emph{Proof:}
We first prove the lower bound. 
Define $E=\{v\in \mathbb{F}_2^n |H(v)=1\}$, where $H(v)$ denotes the Hamming weight
 of $v$. Since the dimension of $U$ is $k$,
 there exists $E'$, a subset of $E$, with $n-k$ elements such that 
\begin{align}\label{Eq:osum}
&U\oplus U'=\mathbb{F}_2^n\nonumber\\
&U\cap U'=\{0\},
\end{align}
 where $U'=\text{span}(E')$ and $\oplus$ denotes the direct sum of two sub-spaces.
 For each $u'\in U'$, define $U_{u'}=U+u'$. Since $U\cap U'=\{0\}$, 
  $U_{u'}$s are disjoint for $u' \in U'$. %\cap U_{u'_2}={\O}$ for $u'_1\neq u_2'$.
   Next, we shall bound $P(U_{u'})$. Suppose $H(u')=r$, then we have
\begin{align}\label{Eq:boundUUp}
&\mathbb{P}[U_{u'}]=\sum_{u\in U_{u'}}\mathbb{P}[u]=\sum_{u\in U}\mathbb{P}[u+u']\nonumber\\
&\le\sum_{u\in U}\mathbb{P}[u]\left(\frac{\max\{p,1-p\}}{\min\{p,1-p\}}\right)^r \nonumber\\
&=\mathbb{P}[U]\left(\frac{\max\{p,1-p\}}{\min\{p,1-p\}}\right)^r, 
\end{align}
%Note that $U_{u'}$s are disjoint and for each $u'\in U'$
where the inequality holds because adding $u'$ to $u$ flips $r$ of 
the coordinates of $u$. Since the elements of $E'$ have Hamming weight $1$
 and $U'=\text{span}(E')$, we have $H(u')\le n-k$ for any $u'\in U'$. Thus, the following holds
\begin{align}\label{Eq:prooflinearalgebra}
1&=\mathbb{P}[\mathbb{F}_2^n]\stackrel{\eqref{Eq:osum}}{=} \mathbb{P}[\cup_{{u'}\in U'}U_{u'}]\nonumber\\
&\stackrel{U_{u'}\text{s are disjoint}}{=}\sum_{u'\in U'}\mathbb{P}[U_{u'}]\nonumber\\
\stackrel{\eqref{Eq:boundUUp}}{\le} &\sum_{u'\in U'}\mathbb{P}[U]\left(\frac{\max\{p,1-p\}}{\min\{p,1-p\}}\right)^r\nonumber\\
&=\mathbb{P}[U]\sum_{r=0}^{n-k}\binom{n-k}{r}\left(\frac{\max\{p,1-p\}}{\min\{p,1-p\}}\right)^r\nonumber\\
&=\mathbb{P}[U]\left(1+\frac{\max\{p,1-p\}}{\min\{p,1-p\}} \right)^{n-k}\nonumber\\
&=\mathbb{P}[U]\left(\frac{1}{\min\{p,1-p\}} \right)^{n-k} .
\end{align}
This shows that,
\begin{align}
\mathbb{P}[U] \ge (\min\{p, 1-p\})^{n-k}. \nonumber
\end{align}
Modifying \eqref{Eq:boundUUp} to obtain a lower bound on $\mathbb{P}[U_{u'}]$
and modifying the third line of \eqref{Eq:prooflinearalgebra}, the upper bound is proved similarly. \hfill $\square$

 A linear encoder, $f:\mathcal{X}^n\to \mathcal{Y}^k$, where $\mathcal{X}=\mathcal{Y}=\{0, 1\}$ 
 is defined as :
 \\Let $G\in \mathbb{F}_2^{n\times k}$ be the generating matrix of the encoder. 
 $G$ is a mapping from $\{0,1\}^n$ to
 $\{0,1\}^{k}$.
 The encoding is as following
$$x\mapsto xG,$$
where all the operations are over $\mathbb{F}_2$. 
%%%%%%%
\begin{theorem}\label{Th:linearencoder}
Assume $X$ has a Bern$(p)$ distribution and $(n,k,\epsilon,t)$ is a LDSC for this source with a linear encoder.
 If $\epsilon < (\min\{p,1-p\})^t$, then $k \ge n$.
\end{theorem}
\emph{Proof:}
In this proof, all linear spaces are over $\mathbb{F}_2$. Without loss of 
 generality, assume $X_1$ is recovered by $Y_1, ..., Y_t$ and the decoder 
 maps $Y^t=0^t$ to $\hat{X_1}=0$. Consider the induced linear mapping $\pi: \mathcal{X}^n\to \mathcal{Y}^t$.
  Since the dimension of the range of $\pi$ is $n$ and the 
  dimension of the image of $\pi$ is at most $t$, we have
  $\dim(ker(\pi)) \ge n-t$. Note that $0^n\in ker(\pi)$ since $\pi$ is 
  a linear mapping.
   If there exists $x^n\in ker(\pi)$ such that $x_1=1$, then
 half of the vectors in $ker(\pi)$ have $x_1=0$ and half of
  them have $x_1=1$ (because $ker(\pi)$ is a linear space over $\mathbb{F}_2$). Since the decoder maps $0^t$ to $\hat{X}_1=0$, 
 then the vectors in $ker(\pi)$ with $x_1=1$ are erroneous.
% Whenever  will 
 %make an error on the half with $x_1=1$
 Eliminate the first coordinate and consider all the
  vectors in $ker(\pi)$ such that $x_1=1$; they will form a subspace of dimension at least 
  $n-t-1$ in a space of dimension $n-1$. Therefore, using Lemma \ref{Lem:Linearalgebra} we obtain
\begin{align}
&\mathbb{P}[\hat{X^n}\neq X^n]  \ge 
\mathbb{P}[\hat{X_1}\neq X_1]  \nonumber\\
&\ge (\min\{p,1-p\})^{n-1-(n-t-1)}=(\min\{p,1-p\})^{t},
\end{align}
which contradicts $\epsilon < (\min\{p, 1-p\})^t$. 
Therefore, for any $x^n \in ker(\pi)$, $x_1=0$. This means that, if we look at the sub-matrix of $G$
of dimension $n\times t$ consisting of the first $t$ columns, the first row is not in the span of the rest of rows. This implies 
that, in the matrix $G$, the first row is not in the span of the rest of rows.
 If we apply the same argument for any $\hat{X}_i$, we conclude that 
 the rows of the matrix $G$ are independent, resulting in $k\ge n$. \hfill $\square$ 
 \begin{corollary}\label{cor:rateoneforlinearencoder}
For any source $X$ with Bern$(p)$ distribution and any locality $t$, the rate of LDSC with a 
linear encoder is
$R^*(t)=1$.
 \end{corollary}
 \emph{Proof:} Using Theorem \ref{Th:linearencoder}, 
 for any $\epsilon< (\min\{p, 1-p\})^t$, we have 
 $k^*(n, \epsilon, t) \ge n$. Thus, 
 $R^*(\epsilon, t)= \limsup_{n\to \infty} \frac{k^*(n, \epsilon, t)}{n} \ge 1$. 
 Therefore, the rate is $R^*(t) \ge \lim_{\epsilon\to 0} R^*(\epsilon, t) \ge 1$. 
 On the other hand, without using any encoding-decoding, we obtain the rate $1$. 
 This completes the proof.  \hfill $\square$

 Corollary \ref{cor:rateoneforlinearencoder} implies that, with local decoder and linear encoder,
  no compression is possible 
 and the rate of best possible scheme is the same as not 
 using any compression scheme. 
 %
%%%%%%%%%%%%%
\subsection{Linear Decoder}

In this section, we consider a local and linear decoder. 
We show that, for a linear decoder, even without locality assumption,
 the rate of compression is $1$.
  This implies if the decoder is linear, then no compression is possible. 
%%%%  
%%%%%%%%%%%%%%%%%%%
\begin{theorem}\label{Th:lineardecoder}
Let $X$ has a Bern $(p)$ distribution. Assume $(n, k, \epsilon)$ is 
a source coding with linear decoder. We have 
\begin{equation}
k \ge n - \frac{\log (1-\epsilon)}{\log\left( \max\{p,1-p\}\right)}.
\end{equation}
%where $k^*(n,\epsilon)$ is the analogy of $k^*_{ld}(n,\epsilon)$ without any constraint on the decoder.
\end{theorem}
\emph{Proof:} 
%%%%%
.
 Assume $e_1,\dots, e_k$ form the canonical basis of $\{0,1\}^k$.
 Since the decoder is linear, it can only recover $\text{Span}\{g(e_1), ..., g(e_k)\}$
 without error and the rest of the elements of $\{0,1\}^n$ are erroneous. 
 Note that the dimension of $\text{Span}\{g(e_1), ..., g(e_k)\}$ is 
 not greater than $k$. 
  Thus, using Lemma \ref{Lem:Linearalgebra}, we obtain
\begin{align}
&P[g(f(X^n))\neq X^n] \ge 1- \mathbb{P}[Span\{g(e_1), ..., g(e_k)\}] \nonumber\\ & \ge 1- (\max\{p,1-p\})^{n-k}\nonumber.
\end{align}
We also know $P[g(f(X^n))\neq X^n] \le \epsilon$. Therefore,
\begin{align}
k \ge n - \frac{\log (1-\epsilon)}{\log( \max\{p,1-p\})}.\nonumber
\end{align}
This completes the proof. \hfill $\square$

%%%%%%%%%%%%%%%%%
\begin{corollary}
Let $X$ be a Bernoulli $(p)$ source. Also let $f:\mathcal{X}^n\mapsto \{0,1\}^k$ and
 $g:\{0,1\}^k \mapsto \mathcal{X}^n$ be the encoder and linear decoder, respectively.
  For any $t-$local decoder we have
  $$R^* (t)=1.$$
  Moreover, without the locality constraint, the rate is still $1$. 
  \end{corollary}
  \emph{Proof:} Using Theorem \ref{Th:lineardecoder},
   if we take minimum over all choices of codes, we obtain 
$$k^*(n,\epsilon) \ge n - \frac{\log (1-\epsilon)}{\log \left(\max\{p,1-p\}\right)}$$
Thus, $R^*(n, \epsilon) \ge 1- \frac{1}{n} \frac{\log (1-\epsilon)}{\log( \max\{p,1-p\})}$. 
Taking $n\to \infty$, we obtain $R^*(\epsilon)\ge 1$ and $R^*\ge 1$ (where 
$R^*$ denote the rate without any assumption on the locality of decoder). 
Therefore, the rate is $1$, because without using any encoding-decoding we can achieve 
rate $1$. 
%
%%%%%%%%%%%%%%%%%%%%%%%%%%%%
%\subsubsection{Linear Encoder}

\subsection{General Encoder-Decoder}
We focus on the special case of $2-$local decoder with a general encoder ($t=2$). 
\begin{theorem}\label{Th:generalencoderdecoderrate1t2}
Let $X$ be a Bern $(p)$ source and $f:\mathcal{X}^n\mapsto \{0,1\}^k$
 and $g:\{0,1\}^k \mapsto \mathcal{X}^n$ be the general encoder and $t$-local decoder.
  Also, assume a $(n, k, \epsilon, t)$-LDSC for this source.
   For $t=2$, if $\epsilon < (\min\{p, 1-p\})^2$, then $k\ge n$. 
 \end{theorem}
\emph{Proof:}
We prove this by contradiction. Without loss of 
generality assume $p\le \frac{1}{2}$. 
For the sake of contradiction, assume $n > k$. We shall show $\epsilon \ge p^2$.\\
 The claim is that if the code can recover $X_1^{k+1}$ i.i.d. 
 Bernoulli $(p)$ with a $2$-local mapping from $Y_1^k$ on a set with
  probability $p(k)$, then $p(k)\le 1-p^2$. This implies  $\epsilon \ge p^2 $.\\
By induction on $k$ we show $p(k)\le 1-p^2$. For $k=1$, by considering 
all $16$ possible encoder functions ($X^2\to Y_1$), it can be 
seen that $p(1) \le 1-p^2$. Assume $p(k-1)\le 1-p^2$.
 Let $X_1$ be recovered by $Y_1$ and $Y_2$. 
 Without loss of generality assume $g_1(0,0)=0$, 
 where for any $1 \le i \le n$, $g_i$ is a mapping with two inputs, producing $\hat{X_i}$, the
 reproduction of $X_i$. 
 We list all the possible cases:
\begin{enumerate}
\item $g_1(0,1)=0$. If we consider the induced mapping from 
$Y_2^{k}$ to $X_2^{k+1}$, by replacing $0$ with $Y_1$ in all the 
mappings that use $Y_1$ as one of their inputs, we obtain a 
local mapping on a set with maximum probability of $p(k-1)$.
 Similarly, since $g_1(1,1)=g_1(1,0)=1$, if we replace $1$ with $Y_1$, 
 we obtain another local mapping on a set with maximum probability $p(k-1)$. 
 Therefore, $p(k) \le p.p(k-1)+\bar{p}.p(k-1)=p(k-1)\le 1- p^2$.
\item $g_1(1,0)=0$. In this case, replace $0$ with $Y_2$ and construct a 
mapping from $(Y_1, Y_3^{k})$ to $X_2^{k+1}$. Similarly, it can be shown $p(k)\le 1-p^2$.
\item $g_1(1,1)=0$. In this case, replace $Y_1$ by $Y_2$ in all the mappings
 that are using $Y_1$ as one of their inputs. Similarly we obtain $p(k) \le 1-p^2$.
 \item $g_1(1,0)=g_1(0,1)=g_1(1,1)=0$. In this case, $X_1=1$ cannot be decoded correctly. 
 Thus, $p(k) \le 1-p \le 1-p^2$. 
\item $g_1(1,0)=g_1(0,1)=g_1(1,1)=1$. For a binary variable, $Y$, let 
$\bar{Y}$ denote its complement ($\bar{Y}=Y+1$, mode $2$ ). In this case, $\bar{g}_1(Y_1,Y_2)=\bar{Y_1}.\bar{Y_2}$.
 In general, we call $Y_1.Y_2$, $\bar{Y_1}.Y_2$, $Y_1.\bar{Y_2}$, $\bar{Y_1}.
 \bar{Y_2}$, and their complements product forms. Next, we consider this case. 
\end{enumerate}
Note that if only one of the $k+1$ decoding functions is not of the product form,
 then considering that mapping and the above
 argument, by induction we obtain $p(k)\le 1-p^2$.
  Now, assume all the mappings are in the product form. 
  If $Y_i$ is appeared in one of the decoding functions as $X_{i_1}=Y_i.Y_j$
   and in another one as its complement, i.e., $X_{i_2}=\bar{Y_i}.Y_k$, then $X_1=X_2=1$
   cannot be recovered and we have $p(k) \le 1-p^2$.
    Therefore, without loss of generality we assume that all the mappings are of the form $Y_i.Y_j$ and 
    no complement is used.\\
Consider a bipartite graph demonstrating the 
relation between the variables $X_1, \dots, X_{k+1}$ and $Y_1, \dots , Y_k$. 
 On the Y-side of it we have $k$ nodes 
corresponding to $Y_i$s and on X-side of it we have $k+1$ nodes 
corresponding to $X_i$s. The degree of each node on X-side is $2$ indicating 
the variables on the Y-side that are involved in decoding of that node. 
If two nodes on the X-side have the same neighbors on the Y-side, then 
we have $X_{i_1}=Y_iY_j$ and $X_{i_2}=Y_iY_j$. Thus, only 
$X_{i_1}=X_{i_2}$ is recoverable and $p(k) \le 1-p^2$. 
Therefore, there exists nodes such that we have 
  $X_{i_1}=Y_iY_j$, $X_{i_2}=Y_iY_k$, and $X_{i_3}=Y_jY_l$ (note that 
  $l$ might be equal to $k$). If $X_{i_1}=1$,
   then we can find a mapping form $(Y_1^{i-1}, Y_{i+1}^k)$ to $(X_{1}^{i_1}, X_{i_1+1}^{k+1})$
    on a set with maximum probability $p(k-1)$. Also, note that $X_{i_1}=0$, $X_{i_2}=X_{i_3}=1$
     is not possible to recover. Therefore, $p(k) \le p p(k-1)+ {(1-p)}(1- p^2) \le 1-p^2$.
The proof is complete for $t=2$. \hfill $ \square$
\begin{corollary}
Let $X$ be a Bernoulli $(p)$ source. Using 
a general encoder and $2-$local decoder, we have % Also let $f:X^n\mapsto \{0,1\}^k$ and
   $$R^* (t)=1.$$
  \end{corollary}
  \emph{Proof:} It follows from Theorem \ref{Th:generalencoderdecoderrate1t2} .
%\\This theorem shows for a general encoder- decoder with locality, $t=2$, the rate of LDSC is 
%$1$. 
\subsection{Scaling Number of Queries}
We give an achievability bound on the rate of LDSC 
with logarithmic number of queries with respect to the source block length.
 The number of queries, $t$, can be a growing function of $n$.
In the conventional source coding (not necessarily local) $t(n)$ is a linear function of $n$.
 Therefore, interesting locality regime are the sub-linear type. 
%regime of our interest is the one where $t(n)$ grows sub-linearly with $n$.
In order to establish an achievability bound on LDSC with scaling queries, we use
the following result on the error exponent of source coding. %to establish an achievability bound on the rate. 
This 
approach is motivated by the achievability bound given in \cite{mazumdar2012update}. 
\begin{theorem}(\cite{csiszar2011information})\label{Th:SCerrorexponent}
For a discrete memoryless source with probability measure $\mathbb{P}_X$ and 
a source encoding with rate $R$, we have:
\\ For any $\epsilon > 0, \exists~\ell_{\epsilon}$ such that for any $n \ge 0$ there exists
an encoding-decoding pair $f_n$ and $g_n$ such that
\begin{equation}\label{Eq:errorexponent}
\mathbb{P}[g_n(f_n(X^n))\neq X^n] \le \ell_{\epsilon} 2^{-n(E^*_b(R)-\epsilon)}, 
\end{equation}
where
$$E^*_b(R)=\min_{Q : H(Q)\ge R}D(Q||P).$$
Moreover, this bound is asymptotically tight. 
\end{theorem}
Now, consider the following construction of an encoder-decoder for a source sequence of 
length $n$, where the source has a Bern$(p)$ distribution:
\\Let rate, $R$, be equal to $(1+\delta)H(X)$. Let $X^n$ be a sequence of source symbols.
Divide this sequence into blocks of length $t(n)$ and apply the encoder-decoder
pair, found by Theorem \ref{Th:SCerrorexponent} to each block separately.
Form an encoder-decoder for $X^n$ by concatenating these
 $\frac{n}{t(n)}$ (for the sake of presentation, without loss of generality, we drop ceiling and floor in 
this analysis) pairs of 
encoder-decoder. We now analyze the error of the concatenated source coding.
Using the union bound, we obtain
\begin{align}
&\mathbb{P}[\hat{X}^n\neq X^n] \nonumber\\
& = \mathbb{P}[\cup_{i=1}^{n/t(n)}\{\hat{X}_{(i-1)t(n)+1}^{i t(n)}\neq X_{(i-1)t(n)+1}^{i t(n)}\}]\nonumber\\
&\le \frac{n}{t(n)} \mathbb{P}[\hat{X}^{t(n)}\neq X^{t(n)}].\nonumber
\end{align}
Using, \eqref{Eq:errorexponent} for any $\epsilon$, we obtain
$$\mathbb{P}[\hat{X}^n\neq X^n] \le  \frac{n}{t(n)} \ell_{\epsilon} 2^{-t(n)(E^*_b(R)-\epsilon)}.$$
Since $R > H(X)$, $E^*_b(R)=\Delta > 0$. Thus, we have
$$\mathbb{P}[\hat{X}^n\neq X^n] \le  \frac{n}{t(n)} \ell_{\epsilon} 2^{-t(n)(\Delta-\epsilon)}.$$
Choosing $\epsilon < \Delta$, this bound goes to zero if %and denote $\Delta-\epsilon$ by $\Gamma$. We want to see for which choices of $t(n)$, this bound goes to zero.
$$t(n) > C \log n,$$
for some constant $C$.
Therefore, we have the following result.
\begin{proposition}\label{Pro:scalingldsc}
Let $X$ be a Bern $(p)$ source. Also, let $f:\mathcal{X}^n\mapsto \{0,1\}^k$
 and $g:\{0,1\}^k \mapsto \mathcal{X}^n$ be an encoder and $t(n)$-local decoder. For any $\epsilon$ and $R>H(X)$,
 there exist constants $C$ and $n_0$ such that for any 
 $n > n_0$, there exist a $(n, nR, C \log n, \epsilon)$-LDSC.
  Moreover, for any $R> H(X)$,  there exists a constant $C$ such  
 $$\lim_{\epsilon\to 0} \limsup_{n\to \infty} \frac{1}{n}k^*(n, \epsilon, C \log n)= R.$$
\end{proposition}
Proposition \ref{Pro:scalingldsc} states that, with a relatively small number of queries ($\log n$), the rate of LDSC
approaches the optimal rate $h(p)$.
%%%%%%%%%%%%%%%%%%%%%%
\\Using the result of \cite{csiszar1982linear} on the error exponent 
of source coding with linear encoder, we have the following 
analogy for codes with linear encoder. 
\begin{corollary}\label{Cor:linearencoderlocaldecoderlocalitylog}
Let $X$ be a Bern$(p)$ source. Also, let $f:\mathcal{X}^n\mapsto \{0,1\}^k$
 and $g:\{0,1\}^k \mapsto \mathcal{X}^n$ be a linear encoder
  and $t(n)$-local decoder, respectively. For any $\epsilon>0$ and $R>H(X)$,
 there exist constants  $C$ and $n_0$ such that for any 
 $n > n_0$, there exist a $(n, nR, C \log n, \epsilon)$-LDSC.
  Moreover, for any $R> H(X)$,  there exists a constant $C$ such  
 $$\lim_{\epsilon\to 0} \limsup_{n\to \infty} \frac{1}{n}k^*(n, \epsilon, C \log n)= R,$$
 where the encoder is assumed to be linear.
\end{corollary}
Note that Corollary \ref{cor:rateoneforlinearencoder} shows the rate 
of LDSC with linear encoder 
and constant number of queries is one. However, Corollary 
\ref{Cor:linearencoderlocaldecoderlocalitylog} shows with $O(\log n)$ 
number of queries we can achieve any rate above the entropy rate. 

%%%%%%%%%%%%%%%%%%%%%%%%%%%%%%%%%%%%%%%%%%%%%%%%%%%%%%%%%%%%%%%%%%%%
\section{Locally Decodable Lossy Source Coding (LDLSC)}\label{sec:rateofLDLSC}
%In this section \emph{Locally Decodable Lossy source Coding (LDLSC)} is studied.
 %First, we formally define the problem.
We first define LDLSC and the fundamental limits of it. Then 
we provide achievability bounds on the rate of LDLSC for 
both scaling and constant number of queries. 

 Consider a separable distortion metric defined as
$$d(x^n,\hat{x}^n)=\frac{1}{n}\sum_{i=1}^n d(x_i,\hat{x}_i),$$
where $d:\mathcal{X}\times \hat{\mathcal{X}}: \to \mathbb{R}^+$ is a distortion 
measure. 
\begin{definition}
A $(n, k, d, t)$-LDLSC is a pair containing an encoder, $f:\mathcal{X}^n\mapsto \mathcal{Y}^k$, and a decoder, $g:\mathcal{Y}^k\mapsto \mathcal{X}^n$
, where the decoder is $t$-local and the distortion is bounded, $\mathbb{E}[d(X^n,g(f(X^n)))]\le d$.\\
Let 
\begin{align}
&k^*(n,d,t)=\nonumber\\
&\min\{k \text{~such that~} \exists (n,k,d,t)- \text{~ LDLSC~}\},
\end{align}
and 
\begin{equation}
R^*(d,t)=\limsup_{n\to\infty} \frac{k^*(n,d,t)}{n}.
\end{equation}
\end{definition}

\begin{note}
We assume a binary source, $\mathbb{F}=\mathbb{F}_2$ with $d(x,\hat{x})=\mathbf{1}\{x\neq\hat{x}\}$.
 In this case we have
$$\mathbb{E}[d(X^n,g(f(X^n)))]=\frac{1}{n}\sum_{i=1}^n \mathbb{P}[X_i \neq \hat{X}_i] \le d,$$
 which is the same as assuming the bit error rate is bounded 
 (comparing to block error rate in the definition of LDSC).
\end{note}

\subsection{Scaling Number of Queries}
In this section we consider the scaling number of queries.
 Therefore, let $t(n)$ be a growing function
of $n$. 
The following is an achievability bound on the rate for 
finite block length.

 %%%%%%%%%%%%%%%%

%%%%%%%%%%%%%%%%
\begin{theorem}\label{Th:ratedistortionfinite}
For a Bern$(p)$ source, a distortion level $d$,
 and any growing number of queries $t(n)$, we have
\begin{equation}\label{eq:ratedistortionfinitegent}
R^*(n,d,t(n)) \le  h(p)-h(d)+\frac{\log t(n)}{t(n)}+o\left(\frac{\log t(n)}{t(n)}\right)
\end{equation}
\end{theorem}
\emph{Proof:}
Recall the finite block length results on source coding \cite{finitesourcecoding}:
%%%%%%%%%%%%%%%%
 For a Bern$(p)$ source, and distortion level $d$, there exists a code such that
\begin{equation}\label{Eq:finitelengthlsc}
 R(n,d) \le h(p)-h(d)+\frac{\log n}{n}+o(\frac{\log n}{n}).
 \end{equation}
 Now, divide the sequence $X^n$ into $\frac{n}{t(n)}$ blocks of length
 $t(n)$ (for the sake of presentation, we drop ceiling and floor in this 
 argument). Apply the encoder-decoder obtained from \eqref{Eq:finitelengthlsc}
 to each block. Concatenate these $\frac{n}{t(n)}$ pairs to obtain
 an encoder-decoder for $X^n$. The average distortion of the 
 overall code is also bounded by $d$, and its rate is bounded by
 $$h(p)-h(d)+\frac{\log t(n)}{t(n)}+o\left(\frac{\log t(n)}{t(n)}\right).$$
 Which concludes the theorem. \hfill $\square$
 %%%%%%%%%%%%%%%%
 \\Theorem \ref{Th:ratedistortionfinite} shows that for any number of queries
  such as $t(n)$, if $\lim_{n\to \infty}t(n)=\infty$, then $R^*(t)=h(p)-h(d)$, 
  which is the rate distortion.
%%%%%%%%%%%%%%%
\begin{corollary}\label{Cor:ratedistortionlogn}
For the special case of $t(n)=t \log n$, we have
\begin{align}\label{Eq:ratedistortionfinite}
&R^*(n, d ,t\log n) \le\nonumber\\
&  h(p)-h(d)+\frac{\log( t\log n)}{t\log n}+o\left(\frac{\log( t\log n)}{t\log n}\right)
\end{align}
\end{corollary}
\emph{Proof:} result of Theorem \ref{Th:ratedistortionfinite} for $t(n)=t\log n$. \hfill $\square$
%%%%%%%%%%%%%%%%%%%%%%%%%%
\\Reference \cite {Mihai} studies the problem of storage of bits with local recovery
 (with the same definition of locality we use here). Those results are based on a generic
transformation of augmented $B$-trees to succinct data structures. They have shown that:
%%%%%%%%%%%%%%%%
 \begin{theorem}[\cite{Mihai}]\label{Mihai}
 Consider a sequence of length $n$ from alphabet $\mathcal{X}$. We can represent this sequence with
 \begin{equation}
 O(|\mathcal{X}|\log n)+n \tilde{H}+\frac{n}{(\frac{\log n}{t})^t}+\bar{O}(n^{3/4})
 \end{equation}
many bits, supporting single bit recovery in $t\log n$ queries, where $\tilde{H}$ denotes the empirical entropy of the 
 sequence.
 Moreover, we can represent a binary sequence of length $u$, with $n$ ones, using 
 \begin{equation}
 \log \binom{u}{n}+\frac{u}{(\frac{\log u}{t})^t}+\bar{O}(u^{3/4})
 \end{equation}
  bits. A decoder exists  querying only $t\log u$ bits to decode 
  any bit of the sequence.
   %Note that the encoder/decoder knows $n$ and $u$ beforehand. 
 \end{theorem}
%%%%%%%%%%%%%%%%

%%%%%%%%%%%%%%%%

 We now compare the bound given in corollary \ref{Cor:ratedistortionlogn} with
the bound suggested by Theorem \ref{Mihai}.
\\Using Theorem \ref{Mihai} and identity $\log \binom{n}{pn}=nh(p)+O(\log n)$, for any $d$, we obtain
\begin{align}\label{Eq:Mihaiach}
&R^*(n,d,t\log n)\le\nonumber\\
& h(p)+O(\frac{\log n}{n})+\frac{1}{(\frac{\log n}{t})^t}+\frac{1}{n}\bar{O}( n^{3/4}).
\end{align}
It is clear that for any fixed $d$, the bound given by \eqref{Eq:ratedistortionfinite} is asymptotically (in $n$) better than \eqref{Eq:Mihaiach}. 
Note that the bound given in \eqref{Eq:Mihaiach} does not gain from the fact that 
encoding-decoding scheme  
can tolerate a distortion $d$.
 One may consider the case where $d$ goes to zero as $n$ goes to infinity.
  Assume both bounds hold for this case as well. We show that if $d(n)=O(\frac{1}{\log n})$,
   then \eqref{Eq:Mihaiach} is tighter than \eqref{Eq:ratedistortionfinite}.
    We omit the last term in both bounds and assume $t=1$. In order to show that,
    \eqref{Eq:Mihaiach} is tighter than \eqref{Eq:ratedistortionfinite}, we need to prove
   \begin{equation}
h(p)+O(\frac{\log n}{n})+\frac{1}{\log n}\le  h(p)-h(d(n))+\frac{\log (\log n)}{\log n}.\nonumber
\end{equation}
   This inequality holds if 
   $$h(d(n))\le \frac{\log (\log n)}{\log n}-O(\frac{\log n}{n})-\frac{1}{\log n}.$$
   It can be seen that for $d(n)=O(\frac{1}{\log n})$, this inequality holds.

%%%%%%%%%%%%%%%%%%%%
\subsection{Constant Number of Queries}
For a given number of queries, we show that one 
can achieve any rate above the rate distortion function with a properly large locality.
Consider the following construction:

%Assume that the source sequence is the i.i.d. product of $X$ with 
%probability measure $\mathbb{P}_X$ ($\mathbb{P}_{X^n}=\mathbb{P}_X^n$).
   For a given $\delta$, we wish to show there exists $t$, such that a LDLSC with
   locality $t$ achieves the rate $(1+\delta)(R(d))$ with average distortion bounded by $d$.
  From \eqref{Eq:finitelengthlsc}, we can get the 
  bound $R(t, d) \le R(d)+2\frac{\log t}{t}$ for large enough $t$. Also,  
  let $t$ be large enough such that $2\frac{\log t}{t} \le \delta R(d)$. Therefore,
   there exists $t$ such that 
   $$R(t,d) \le R(d)(1+\delta).$$
  Thus, there exists an encoder and decoder pair for $X^t$, such that the
  rate of the code is less than $(1+\delta)R(d)$ and the distortion is bounded by $d$. 
  Now, consider $n$ pairs of the same encoder-decoder. Concatenate these encoder-decoder pairs to
  form an encoder-decoder for $X^{nt}$. In this way, we obtain a source coding for $X^{nt}$ with 
  distortion 
  \begin{align}
&  \mathbb{E}[\frac{1}{nt}\sum_{i=1}^{nt} d(x_i,\hat{x_i})]\nonumber\\
  &=\frac{1}{n}\sum_{j=1}^n \mathbb{E}[d(X_{(j-1)t+1}^{jt},\hat{X}_{(j-1)t+1}^{jt})] \le d,
  \end{align}
  and rate 
  $$R^*(nt,d, t)\le R(d)(1+\delta).$$
  Therefore, for any block length, 
  there exists a $t$-local LDLSC with rate $(1+\delta)R(d)$ and average distortion bounded by $d$ for this source.
  \begin{proposition}
  For any source $X$ with probability measure $\mathbb{P}_X$ and any distortion measure, 
  and distortion level, $d$, $R(d)$ is
  $$\inf\{ R~:~\exists~ t \text{ and a sequence of }t- \text{ LDLSC with rate } R \}.$$
%  Moreover, for a rate equal to $R(d)(1+\delta)$, it is enough to
  %have $\frac{t}{\log t}  \ge  \frac{1}{\delta R(d)}$.
  \end{proposition}
This proposition states that, in order to achieve the rate $(1+\delta)R(d)$, one need to choose $t$ to be 
roughly $\frac{1}{\delta R(d)}$.

%%%%%%%%%%%%%%%%%%%%%%%%%%%%%%%%%%%%%%%%%%%%%%%%%%%%%%%%%%%%%%%%%%%%
\section{Conclusion and Future Work}\label{sec:conclusion}
We introduced locally decodable source coding in both almost lossless and 
lossy cases. The following summarizes the main results we showed in this work:
\begin{itemize}
\item Almost lossless source coding:
\begin{itemize}
\item Constant locality: We show that, the rate of linear LDSC is one, meaning that
no compression is possible. Moreover, we show that, the rate of source coding with 
a general encoder and a linear decoder (not necessarily local) is one, meaning that 
no compression is possible. %linear decoder 
Also for locality, $t=2$, the rate of any encoder-decoder 
is one. A future work is to consider LDSC with a general encoder and $t-$ local decoder ($t > 3$) and study 
the converses bounds on it.
\item Scaling locality: We can achieve any given rate above the Shannon fundamental entropy rate, with
 logarithmic locality in the block-length.
\end{itemize}
\item Lossy source coding: \begin{itemize}
\item Constant locality: Any given rate above the Shannon fundamental rate distortion is achievable with
 a proper constant locality. This locality is proportional to the inverse of the difference between the given rate and rate distortion.
\item Scaling locality: Shannon fundamental rate distortion is achievable with
 any scaling locality ($\lim_{n\to\infty} t(n)=\infty$) and the rate of convergence is upper bounded as in Theorem \ref{Th:ratedistortionfinite}. 
 We show that, this upper bound is asymptotically tighter than the existing bounds in data structure literature. 
\end{itemize}
\end{itemize}
%We also studied fixed to variable length local encoder-decoder. 
%%%%%%%%%%%%%%%%%%%%%%%%%%%%%%%%%%%%%%%%%%%%%%%%%%%%%%%%%%%%%%%%%%%%
\bibliographystyle{IEEEtran}
\bibliography{references}

\newcommand{\noopsort}[1]{}
\begin{thebibliography}{10}
\providecommand{\url}[1]{#1}
\csname url@samestyle\endcsname
\providecommand{\newblock}{\relax}
\providecommand{\bibinfo}[2]{#2}
\providecommand{\BIBentrySTDinterwordspacing}{\spaceskip=0pt\relax}
\providecommand{\BIBentryALTinterwordstretchfactor}{4}
\providecommand{\BIBentryALTinterwordspacing}{\spaceskip=\fontdimen2\font plus
\BIBentryALTinterwordstretchfactor\fontdimen3\font minus
  \fontdimen4\font\relax}
\providecommand{\BIBforeignlanguage}[2]{{%
\expandafter\ifx\csname l@#1\endcsname\relax
\typeout{** WARNING: IEEEtran.bst: No hyphenation pattern has been}%
\typeout{** loaded for the language `#1'. Using the pattern for}%
\typeout{** the default language instead.}%
\else
\language=\csname l@#1\endcsname
\fi
#2}}
\providecommand{\BIBdecl}{\relax}
\BIBdecl

\bibitem{shannon1948mathematical}
C.~E. Shannon, ``A mathematical theory of communication,'' \emph{Bell Syst.
  Tech. J.}, vol.~27, pp. 379--423, 1948.

\bibitem{mackay1999good}
D.~MacKay, ``Good error-correcting codes based on very sparse matrices,''
  \emph{Information Theory, IEEE Transactions on}, vol.~45, no.~2, pp.
  399--431, 1999.

\bibitem{Mihai}
M.~Patrascu, ``Succincter,'' in \emph{Foundations of Computer Science, 2008.
  FOCS'08. IEEE 49th Annual IEEE Symposium on}.\hskip 1em plus 0.5em minus
  0.4em\relax IEEE, 2008, pp. 305--313.

\bibitem{bloom1970space}
B.~H. Bloom, ``Space/time trade-offs in hash coding with allowable errors,''
  \emph{Communications of the ACM}, vol.~13, no.~7, pp. 422--426, 1970.

\bibitem{pagh2001low}
R.~Pagh, ``Low redundancy in static dictionaries with constant query time,''
  \emph{SIAM Journal on Computing}, vol.~31, no.~2, pp. 353--363, 2001.

\bibitem{jacobson1989space}
G.~Jacobson, ``Space-efficient static trees and graphs,'' in \emph{Foundations
  of Computer Science, 1989., 30th Annual Symposium on}.\hskip 1em plus 0.5em
  minus 0.4em\relax IEEE, 1989, pp. 549--554.

\bibitem{chandar2009locally}
V.~Chandar, D.~Shah, and G.~Wornell, ``A locally encodable and decodable
  compressed data structure,'' in \emph{Communication, Control, and Computing,
  2009. Allerton 2009. 47th Annual Allerton Conference on}.\hskip 1em plus
  0.5em minus 0.4em\relax IEEE, 2009, pp. 613--619.

\bibitem{neuhoff1982causal}
D.~Neuhoff and R.~Gilbert, ``Causal source codes,'' \emph{Information Theory,
  IEEE Transactions on}, vol.~28, no.~5, pp. 701--713, 1982.

\bibitem{DBLP:journals/corr/abs-1301-0079}
Y.~Kaspi and N.~Merhav, ``Zero-delay and causal single-user and multi-user
  lossy source coding with decoder side information,'' \emph{CoRR}, vol.
  abs/1301.0079, 2013.

\bibitem{yekhanin2006new}
S.~Yekhanin, ``New locally decodable codes and private information retrieval
  schemes,'' in \emph{Electronic Colloquium on Computational Complexity, vol.
  TR06}, 2006, p. 127.

\bibitem{papailiopoulos2012locally}
D.~Papailiopoulos and A.~Dimakis, ``Locally repairable codes,'' in
  \emph{Information Theory Proceedings (ISIT), 2012 IEEE International
  Symposium on}.\hskip 1em plus 0.5em minus 0.4em\relax IEEE, 2012, pp.
  2771--2775.

\bibitem{varshney2008malleable}
L.~R. Varshney, J.~Kusuma, and V.~K. Goyal, ``Malleable coding: Compressed
  palimpsests,'' \emph{arXiv preprint arXiv:0806.4722}, 2008.

\bibitem{montanari2008smooth}
A.~Montanari and E.~Mossel, ``Smooth compression, gallager bound and nonlinear
  sparse-graph codes,'' in \emph{Information Theory, 2008. ISIT 2008. IEEE
  International Symposium on}.\hskip 1em plus 0.5em minus 0.4em\relax IEEE,
  2008, pp. 2474--2478.

\bibitem{mazumdar2012update}
A.~Mazumdar, G.~W. Wornell, and V.~Chandar, ``Update efficient codes for error
  correction,'' in \emph{Information Theory Proceedings (ISIT), 2012 IEEE
  International Symposium on}.\hskip 1em plus 0.5em minus 0.4em\relax IEEE,
  2012, pp. 1558--1562.

\bibitem{csiszar2011information}
I.~Csiszar and J.~K{\"o}rner, \emph{Information theory: Coding theorems for
  discrete memoryless systems}.\hskip 1em plus 0.5em minus 0.4em\relax
  Cambridge University Press, 2011.

\bibitem{csiszar1982linear}
I.~Csiszar, ``Linear codes for sources and source networks: Error exponents,
  universal coding,'' \emph{Information Theory, IEEE Transactions on}, vol.~28,
  no.~4, pp. 585--592, 1982.

\bibitem{finitesourcecoding}
Z.~Zhang, E.~Yang, and V.~Wei, ``The redundancy of source coding with a
  fidelity criterion. 1. known statistics,'' \emph{Information Theory, IEEE
  Transactions on}, vol.~43, no.~1, pp. 71--91, 1997.

\end{thebibliography}

\end{document}